\documentclass[journal]{IEEEtran}
\ifCLASSINFOpdf
  \usepackage[pdftex]{graphicx}
  \usepackage{caption}
  \usepackage{booktabs, tabularx, makecell}
  \usepackage{cite}
\else
\fi
\begin{document}
%
\title{Multimodal Neuroimaging Attention-Based architecture for Cognitive Decline Prediction}
%
%
%

\author{Jamie Vo, Naeha Sharif and Ghulam Mubashar Hassan \\ Department of Computer Science \& Software Engineering, University of Western Australia
\\ 22735819@student.uwa.edu.au, naeha.sharif@uwa.edu.au, ghulam.hassan@uwa.edu.au}

%
%

\maketitle

\begin{abstract}
The early detection of Alzheimer's Disease is imperative to ensure early treatment and improve patient outcomes.
There has consequently been extenstive research into detecting
AD and its intermediate phase, mild cognitive impairment
(MCI). However, there is very small literature in predicting the conversion to AD and MCI from
normal cognitive condition. Recently, multiple studies have applied
convolutional neural networks (CNN) 
which integrate Magnetic Resonance Imaging (MRI) and
Positron Emission Tomography (PET) to classify
MCI and AD. However, in these works, the fusion of MRI
and PET features are simply achieved through concatenation,
resulting in a lack of cross-modal interactions. In this paper, we
propose a novel multimodal neuroimaging attention-based CNN
architecture, MNA-net, to predict whether cognitively normal
(CN) individuals will develop MCI or AD within a period of 10
years. To address the lack of interactions across neuroimaging
modalities seen in previous works, MNA-net utilises attention
mechanisms to form shared representations of the MRI and PET
images. The proposed MNA-net is tested in OASIS-3 dataset
and is able to predict CN individuals who converted to MCI
or AD with an accuracy of 83\%, true negative rate of 80\%,
and true positive rate of 86\%. The new state of the art results
improved by 5\% and 10\% for accuracy and true negative rate
by the use of attention mechanism. These results demonstrate the
potential of the proposed model to predict cognitive impairment
and attention based mechanisms in the fusion of different
neuroimaging modalities to improve the prediction of cognitive
decline.

\end{abstract}


%
\IEEEpeerreviewmaketitle

\section{Introduction}
%
%
%
%
\IEEEPARstart{A}{lzheimer’s} disease (AD) is a neurodegenerative condition characterised by memory impairment and cognitive decline, eventually progressing to permanent neuron injury and death \cite{jagust2013vulnerable,salami2022}. As of now, there is no cure for AD, leaving current treatment processes to revolve around delaying the onset of cognitive symptoms \cite{sperling2011toward}. There has consequently been a larger focus placed on improving the diagnostic methods for the early detection of AD, as well as its intermediate phase, mild cognitive impairment (MCI), to ensure early treatment.

The clinical inspection of neuroimages such as Magnetic Resonance Imaging (MRI) and Positron Emission Tomography (PET) is currently performed extensively in the diagnosis of AD~\cite{johnson2012brain,roman2012contribution,petrella2003neuroimaging}. They provide the ability to detect common biomarkers related to AD such as atrophy in brain regions and the presence of amyloid plaques. However, despite the use of neuroimages, the diagnosis of AD is still difficult due to the lack of clarity into the characteristics relating to the pathogenesis of the disease \cite{jagust2013vulnerable}.

With the advancement in artificial intelligence (AI) and especially deep learning (DL) algorithms, there has been a rapid growth in the applications of neural networks in computer aided diagnosis (CAD) systems for MCI and AD diagnosis. Deep learning and CAD systems provide a more sensitive diagnostic method capable of identifying the underlying characteristics relating to the pathogenesis of MCI and AD which may normally be undetected by humans. In recent works, convolutional neural networks (CNN) have been implemented in multimodal ensemble networks which combine multiple neuroimaging modalities such as MRI and PET \cite{tu2022alzheimer,liu2018multi}. By leveraging information from multiple modalities, these models were able to have a more comprehensive understanding of the pathogenesis of MCI and AD, resulting in greater performance when compared against models trained on single neuroimaging modalities.

Although the performance of multimodal neuroimaging models have been shown to be superior in the classification of MCI and AD, they are limited by their lack of cross-modal interactions. Different neuroimaging modalities may have varying relationships and influences with each other. For example, certain regions in an MRI image could possibly enhance or complement certain features in a PET image and vice versa. In previous works, learnt features of the different neuroimaging modalities are simply concatenated, limiting the ability of the models to learn relationships and form shared representations \cite{odusami2023,velazquez2022multimodal}. Additionally, there is a lack of work in the use of multimodal models in predicting the conversion of cognitively normal (CN) individuals to cognitive decline. As early treatment is imperative for patient outcomes, it is beneficial to predict whether an individual will develop MCI or AD in the future.

In this work, we present a multimodal neuroimaging attention-based CNN, MNA-net, to address the aforementioned points. MNA-net consists of a patch based architecture and utilises multi-headed self-attention mechanisms to combine both MRI and PET features to predict the conversion of CN individuals to MCI or AD. We propose that the use of attention mechanisms to create shared representations of MRI and PET features provides much more meaningful information to aid in the prediction of cognitive decline and therefore improve model performance.

In summary, this paper's main contributions are: 
\begin{itemize}

  \item {To develop a multimodal model to detect the progression of cognitive decline in CN individuals.}

  \item {To evaluate the performance of attention-based mechanisms for the fusion of PET and MRI features in the prediction of cognitive decline.}

\end{itemize}

\section{Related Works}
Earlier works into CAD systems for the diagnosis of MCI and AD typically utilised traditional machine learning techniques. Rezaei \textit{et al.} \cite{trambaiolli2010support} proposed a support vector machine (SVM) classifier trained on MRI images separated into region of interests (ROI) for each subject. They were able to achieve an accuracy of 88.34\% when classifying CN individuals against individuals with AD. Similarly,  Vaitinathan \textit{et al.} \cite{vaithinathan2019novel} used a region-based method to extract features from MRI images for the classification of AD. The proposed method involved extracting texture features from image blocks based on defined ROIs in the 2D MRI slices. Using the texture features, binary classification was then performed between CN, MCI, and AD patients using a linear SVM, random forest classifier, and a K-nearest Neighbours classifier. For AD vs CN classification, the random forest and K-nearest Neighbours classifiers performed the best, both achieving accuracies of 87.39.\%

While traditional machine learning approaches have seen success in the classification of MCI and AD, to effectively analyse and detect patterns in neuroimages, hand-crafted features and feature extraction methods are necessary. These are often very complex and require domain and clinical expertise. Consequently, there has been growing interest in developing CAD systems using deep learning algorithms that will automatically learn their own features for the classification of MCI and AD. Gunawaderna \textit{et al.} \cite{gunawardena2017applying} proposed a 2D CNN trained on 2D MRI image slices for the classification of AD, MCI, and CN individuals. Their model was able to achieve an accuracy, sensitivity, and specificity of 96\%, 96\%, and 98\% respectively. Tufail \textit{et al.} \cite{tufail2020multiclass} however, found that the 3D CNNs outperformed their 2D CNN counterparts. They trained two 2D CNNs and 3D CNNs, one of the 2D and 3D CNNs of which were trained on MRI images, and the others trained on PET images. They found that both 3D CNNs outperformed their 2D counterparts, with the 3D CNN trained on PET images performing the best. By using 3D convolutions in the networks, there is no loss in spatial information, allowing for improved model performance. 

To this date, as per our literature survey, there has only been one study which has employed CNNs to predict the conversion of CN individuals to MCI or AD. Bardwell \textit{et al.} \cite{bardwell2022cognitive} proposed a 3D CNN utilising a patch-based approach. MRI volumes were divided into 27 uniform patches and fed into individual 3D CNNs. Learnt features from each CNNs were then concatenated and passed through a logistic regression model for binary classification. They were able to achieve an accuracy of 90\% when predicting whether a CN individual would develop MCI or AD within a period of 3000 days. However, the study was tested on a limited size dataset.

In more recent works, there has been a growing number of applications of ensemble-based architectures that allow researchers to leverage the strengths from multiple modalities for MCI and AD CAD systems. The pathogenesis of Alzheimer’s is complex \cite{jagust2013vulnerable} and therefore multiple image modalities, clinical data, and cognitive assessments may be required to effectively detect the presence or development of MCI and AD in individuals. Velazquez and Lee \cite{velazquez2022multimodal} proposed an ensemble model consisting of a random forest classifier and a CNN to predict the conversion to AD from MCI. The random forest model was trained on patient biometric and clinical test scores, while the CNN was trained on diffusion tensor imaging (DTI) scans. They were able to achieve an accuracy of 98.81\% for MCI to AD conversion prediction.

Feng \textit{et al.} \cite{feng2019deep} proposed a deep learning based framework for AD classification utilising both MRI and PET volumes. They trained two 3D CNNs to first extract the intrinsic features of the MRI and PET volumes. A Fully Stacked Bi-directional Long Short Term Memory (FSBi-LSTM) architecture was then adopted to learn the spatial information in neuroimages using the 3D CNN outputs. Their proposed architecture was able to achieve an accuracy of 94.82\% in AD vs CN classification. 

While there are promising results in the use of multimodal neuroimages for AD classification, many studies are limited by the lack of cross-modal interactions. The different neuroimaging modalities may have complex relationships and provide complementary information to each other that could aid in the classification of MCI and AD. In many of the current studies, learned features of the different modalities are simply concatenated, limiting the model's ability to learn the relationships between the different modalities and form shared representations. In an attempt to address a similar problem when combining non-imaging and imaging data, Golovanevsky \textit{et al.} \cite{golovanevsky2022multimodal} proposed a multimodal attention-based architecture for AD diagnosis. Their study utilised three different modalities: genetic data, clinical data, such as memory tests and subject demographics, and MRI volumes. In their proposed architecture, a 3D CNN was trained using the 3D MRI volumes, while two individual deep neural networks were trained using the genetic and clinical data. Outputs of the three models were then each passed through a self-attention layer followed by a cross-modal attention layer to create new shared representations of each modality that take into account the other modalities. Outputs of all the cross-modal attention layers are then finally concatenated and fed through a fully connected layer for classification. Their proposed model was able to achieve an accuracy of 96.88\% in the classification of CN, MCI, and AD individuals. 

Inspired by the literature, this study seeks to investigate the effect of attention-based mechanisms in combining PET and MRI image features to predict the conversion of CN individuals to MCI or AD.

\section{Materials and Methods}
\subsection{Data Collection}
All data used in this study was obtained from Open Access Series of Imaging Studies (OASIS), OASIS-3 dataset \cite{lamontagne2019oasis}. OASIS was launched in 2007 with the primary goal of making neuroimaging data publicly available for study and analysis. OASIS-3 is a longitudinal dataset released as a part of OASIS in 2018. It is a compilation of clinical data and MRI and PET images of multiple subjects at various stages of cognitive decline collected over the course of 30 years. Subject cognitive states in OASIS-3 are defined by Clinical Dementia Rating (CDR) scores. A total of 1378 participants entered the study, 755 of which were cognitively normal (CDR = 0), and 622 who were at progressing stages of cognitive decline (CDR \(\geq 0.5\)). For our study,  the MRI and PIB PET images in OASIS-3 were used.

\subsection{Subject Selection}
To predict the progression of cognitive impairment in individuals within OASIS-3, two groups of subjects were of interest: subjects who remained CN, and subjects transitioned from CN to MCI or AD over the course of the study in OASIS-3. For this scope of this work, a timeframe 10 years was considered. A key factor to consider during subject selection is the temporal alignment of data. It is important that subject scans are taken within close proximity of their initial diagnosis to ensure that scans are representative of their cognition at the time of their baseline. Taking these factors into consideration, the subject selection criteria for the OASIS-3 dataset were as follows:

\begin{enumerate}
\item Subjects were diagnosed as CN at baseline.
\item Subjects have taken MRI and PET scans that are within a year from their baseline diagnosis. 
\item Of CN subjects who developed cognitive impairment over the course of the study, only those who were diagnosed with MCI or AD within 10 years of their baseline diagnosis were considered. 
\item Of subjects who remained CN over the course of the study, only those who received a diagnosis of CN at least 10 years after their baseline diagnosis were considered. 

\end{enumerate}
After applying the selection criteria, 204 subjects remained in the final dataset. Of these subjects, 104 developed some form of cognitive impairment within 10 years, with the remaining 100 subjects remaining CN. A 80/20/20 split was applied for the training, validation, and test sets respectively. Table \ref{tab:1} presents the split of subjects in each class after the subject selection process.
\begin{table}[htbp]
\caption{Subject Class Splits\\}
\label{tab:1}
\centering
\begin{tabular}{cccc}
\toprule 
\textup{Class}       & Training &  Validation & Test\\
\midrule
Remain Cognitively Normal                 & 60 & 20 & 20\\
Develop Cognitive Impairment              & 62 & 21 & 21\\
\bottomrule
\end{tabular}
\end{table}

\subsection{Image Processing}
\begin{figure*}[htbp]
    \centering
    \includegraphics[width=0.55\textwidth]{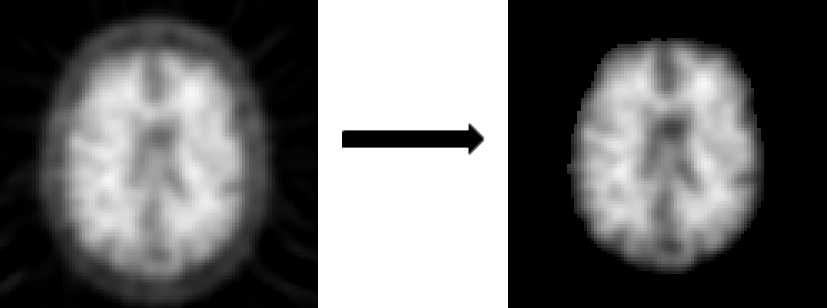}
    \caption{Example of noise and skull removal from a PIB PET image in grayscale.}
    \label{fig:example1}
\end{figure*}

Post processed Freesurfer files for the MRI images were provided by OASIS-3. These files contain the subject-specific 3D MRI images which have undergone skull stripping. PIB PET images, however, were provided as 4D Nifti files. The PET images were acquired in multiple frames over different time intervals post injection of the radiotracer (24 x 5 sec frames; 9 x 20 sec frames; 10 x 1 min frames; 9 x 5 min frames). Temporal averaging of 4D PET images was performed to average the frames into static 3D images. Noise and skull were removed from the PET images using Brain Extration Tool (BET) \cite{jenkinson2005bet2} and Synthstrip \cite{hoopes2022}. Figure \ref{fig:example1} presents an example of PET noise and skull removal. Finally, both MRI and PET images were standardised and aligned to a common anatomical template by normalising voxel intensities and registering them to Montreal Neurological Institute (MNI) space using FMRIB's Linear Image Registration Tool (FLIRT) \cite{jenkinson2002improved}. The final output images were of size 90x116x90.

\begin{figure*}[htbp]
    \centering
    \includegraphics[width=0.75\textwidth]{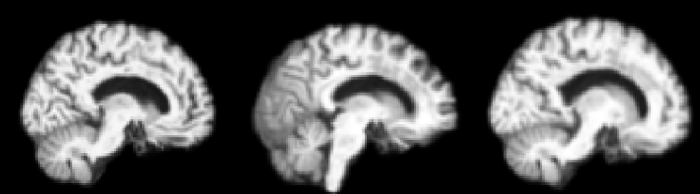}
    \caption{Example of data augmentation applied to an MRI image. From left to right: Normal Brain, Affine Transformed Brain, Elastically Deformed Brain}
    \label{fig:example2}
\end{figure*}

Data augmentation was performed on the training set to increase the dataset size. To simulate different positions and size of the patient within the scanner, and anatomical variations present in the images, random affine transformations and elastic deformations were applied to the images. The resulting training set was of size 488. Figure \ref{fig:example2} presents examples of elastic deformations and affine transforms applied to an MRI image. 

\subsection{Proposed Architecture: MNA-net}
\begin{figure*}[htbp]
    \centering
    \includegraphics[width=0.7\textwidth]{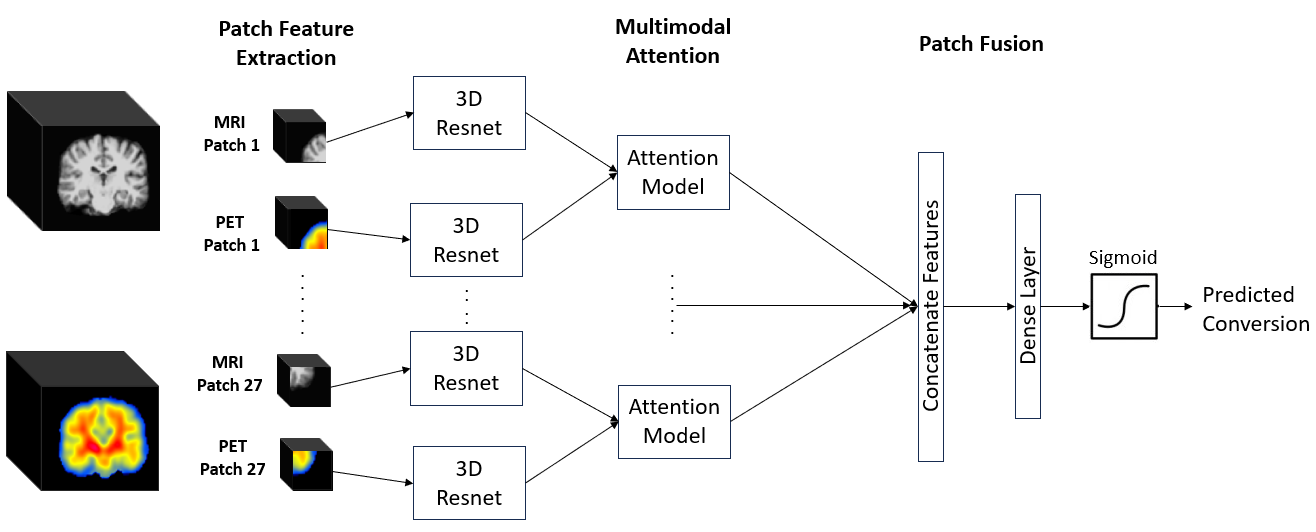}
    \caption{MNA-net architecture}
    \label{fig:MNANet}
\end{figure*}

To harness the strengths of both MRI and PET in the prediction of CN conversion to MCI and AD, we present MNA-net, a multimodal neuroimaging attention-based CNN. We define three stages in the classification process in MNA-net as shown in Figure \ref{fig:MNANet}: patch feature extraction, multimodal attention, and patch fusion. 

In the first stage, we adopt a patch-based technique. MRI and PET images are both divided into 27 uniform patches of size 44 x 54 x 44 with 50\% overlap. Each patch is then fed into a 3D ResNet-10 model to extract the local features of each image. In the second stage of the classification process, we introduce an attention-based ensemble architecture to facilitate the fusion of the different neuroimaging modalities. For every patch in corresponding positions between the MRI and PET patches, we extract the learnt features from the ResNet-10 models and pass them through an attention-based model. This model utilises self-attention mechanisms to enable the model to create shared representations of the MRI and PET features. In the final stage, we consolidate the features extracted from the patch-level models. The attention weighted multimodal features for each patch are extracted from the attention models and flattened, concatenated, and passed through a dense with sigmoid activation for the final classification.

Due to the complexity and wideness of the architecture, training MNA-net as a single model is computationally intensive. Instead, we train the individual models for each classification stage separately. Features are extracted from each model and used as inputs for the subsequent classification stage. 

\subsubsection{Patch-based Feature Extraction}
\begin{figure*}[htbp]
    \centering
    \includegraphics[width=0.7\textwidth]{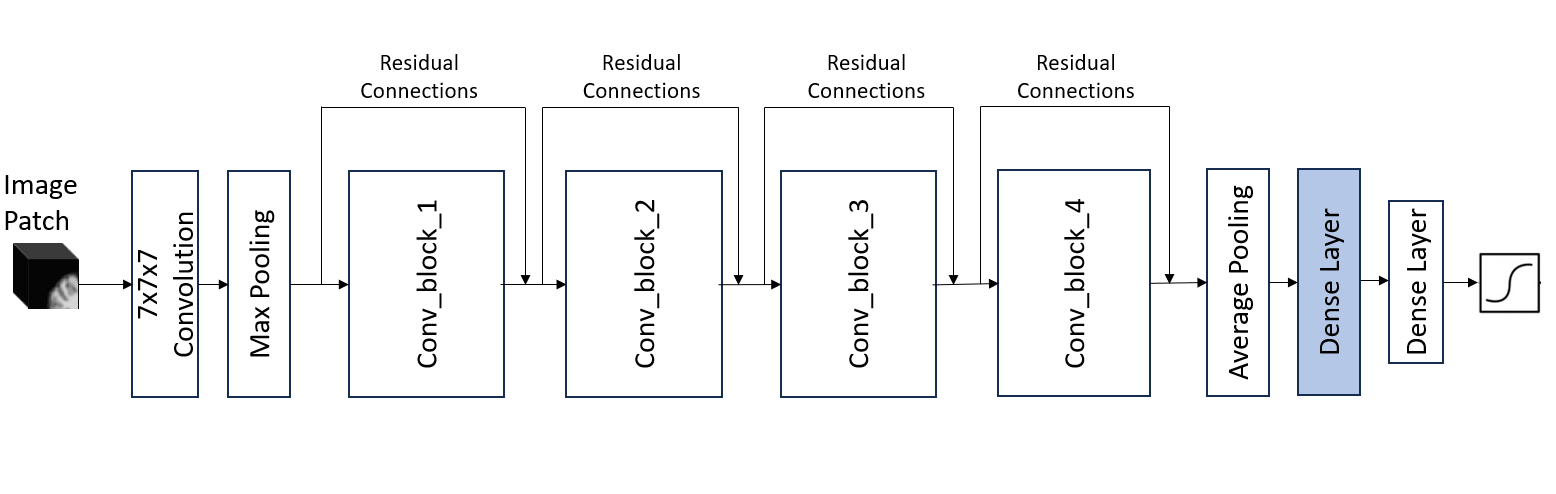}
    \caption{Patch-based Feature Extraction stage: A 3D ResNet-10 architecture where the features from the dense layer prior to the final dense and sigmoid layer (coloured in blue) are extracted and used as inputs for the multimodal attention classification stage.}
    \label{fig:Resnet}
\end{figure*}

To extract the patch-based features, we adopt a 3D ResNet architecture using 3D convolutions adapted from Hara \textit{et al} \cite{hara2017learning} as the backbone model  . A brief illustration of the model is shown in Figure \ref{fig:Resnet}. The patch images of size 44x54x44 are first passed through a 7x7x7 convolutional layer with stride 2 and padding 3, followed by max pooling, batch normalisation, and a ReLu. We then introduce the residual connections through four sequential \textit{conv\_blocks}. Each \textit{conv\_block} consists of two 3x3x3 convolutional layers, each followed by batch normalisation and Relu. A residual connection is included between the beginning of the block and the layer preceding the final ReLu. Strides of 2 are used in the convolutional layers of \textit{conv\_block\_2}, \textit{conv\_block\_3}, and \textit{conv\_block\_4} to perform down sampling. The output feature maps of \textit{conv\_block\_4} are then finally subjected to an average pooling layer, flattened, and subsequently passed through a fully connected layer for final classification. The features prior to the final dense and sigmoid layers are extracted and used as inputs for the multimodal attention classification stage.

\subsubsection{Attention-based Multimodal Feature Fusion}
\begin{figure*}[htbp]
    \centering
    \includegraphics[width=0.7\textwidth]{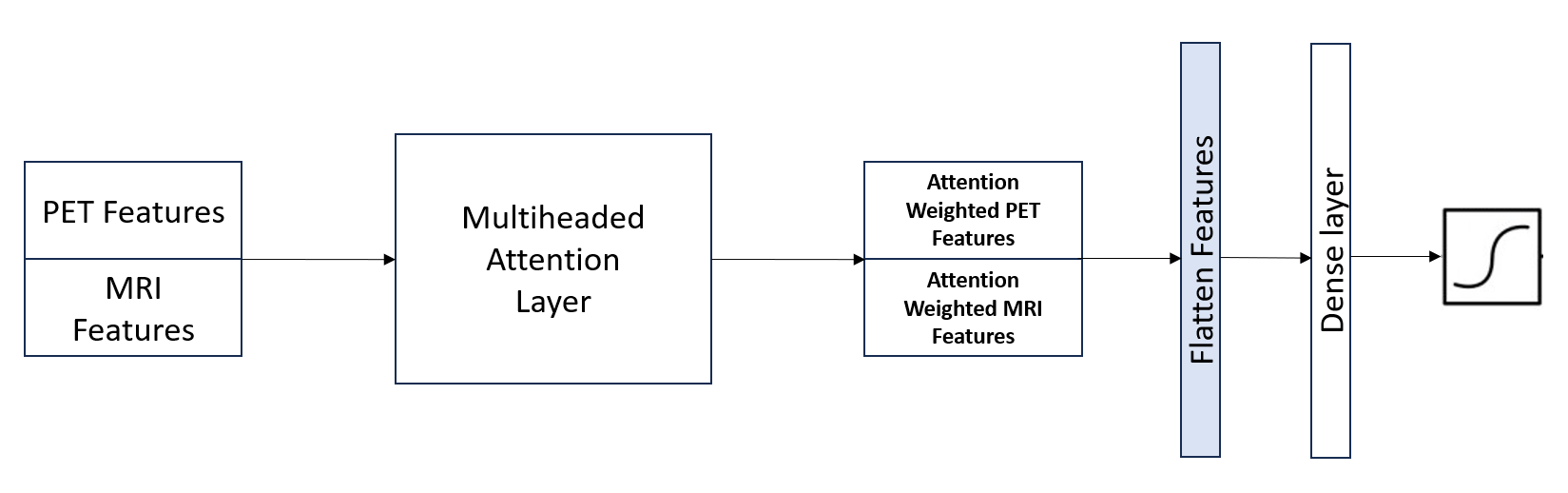}
    \caption{Attention model architecture: The features from the dense layer prior to the final dense and sigmoid layer (coloured in blue) are extracted and used as inputs for the patch fusion classification stage.}
    \label{fig:AttMod}
\end{figure*}

To combine the learned patch features of MRI and PET, we introduce the concept of self-attention into our fusion pipeline. Figure \ref{fig:AttMod} shows the architecture of the attention model trained to fuse the patch features. Multiple approaches of the fusion of PET and MRI for MCI and AD classification found in literature have simply involved the concatenation learned features. This, however, has disadvantages due to the lack of cross modal interactions. Representations of MRI and PET features which take into account information from each other may be more informative than considering each feature independently. Attention mechanisms aim to mimic the cognitive process of attention, enabling neural networks to create shared representations which consider all parts of the input data based on attention scores.

For every patch in corresponding positions between the MRI and PET patches, we extract and vertically stack the features prior to the last layer from the previously trained patch-based feature extraction models. We then pass the stacked features through a multi-head attention layer with 4 attention heads. Finally, the vertically stacked attention weighted outputs for the PET and MRI features are flattened and passed through a fully connected layer for final classification. The final flattened features are then used as inputs to the final model shown in Figure \ref{fig:MNANet}.

The attention mechanism operates on three fundamental components: queries Q, keys K, and values V.  These components are representations of the input data which are learnt during training and used to compute the attention weighted output. Each query in the input data is compared to every key using a similarity function to compute attention weighted scores. These scores determine how much the model should attend to in each value when forming the final representations.

In self-attention, the query, key, and value representations are all derived from the same input during training. In the context of this work, the PET and MRI features will each have a query, key and value representation. To compute the attention weights, we use a scaled dot-product as a similarity function. We compute the dot product of the every query against each key, divide each by the square root of the key dimension, and apply a softmax. Finally, for each query, we multiply the value representations by the computed weights for that query, and sum them to compute the final attention weighted outputs which consider all parts of the input data based on the attention scores. 

In practice, the queries, keys, and values are each represented as matrices. As such, the attention weighted outputs for each query can be computed simultaneously with the following
equation:
\begin{equation}
\mathit{Attention(Q, K, V) = softmax(\frac{QK^{T}}{\sqrt{d_{k}}})V}
\end{equation}
\\where d\textsubscript{k} is the dimension of the key representation.

For this work we use an extension of this concept called multi-headed attention. Instead of computing the scaled dot product attention once, we instead project the queries, keys, and values into H different dimensions using learnt linear projections. Self-attention is then computed for each of the newly projected query, key and values in parallel, concatenated and finally multiplied by a matrix W\textsuperscript{0} to re-project the data. This mechanism allows the model to simultaneously attend to information from different representations and projections of the input data. 

\begin{equation}
\mathit{MultiHead(Q, K, V) = Concat(head_1, ..., head_h)W^{0}}
\end{equation}

\subsection{Experimental Settings and Evaluation Metrics}

The proposed models were implemented using Pytorch library and trained on a 3080 RTX GPU. Each model was trained for 1000 epochs using early stopping with the validation set to prevent overfitting. A learning rate of 0.001 was selected for the Patch-based Feature Extraction and Attention models, while a learning rate of 0.0001 was selected for MNA-net for final classification. Binary cross entropy (BCE) loss and stochastic gradient descent (SGD) with momentum were used for the loss function and optimiser, respectively. A summary of the training parameters for the models is presented in Table \ref{tab:2}.

\begin{table}[htbp]
\caption{Model Training Parameters\\}
\label{tab:2}
\centering
\begin{tabular}{cccc}
\toprule 
\textup{Parameter}       & P-B FE &  Attention Model & MNA-net\\
\midrule
Learning Rate                 & 0.001 & 0.001 & 0.0001\\
Optimiser              & SGD & SGD & SGD\\
Momentum              & 0.9 & 0.9 & 0.9\\
Loss              & BCE & BCE & BCE\\
Batch Size              & 16 & 16 & 10\\
Epochs              & 1000 & 1000 & 1000\\
Early Stopping Patience              & 20 & 20 & 20\\
\bottomrule
\end{tabular}
\raggedright
\\ P-B FE: Patch-based Feature Extraction
\end{table}

To evaluate MNA-net’s performance, we utilise accuracy, true negative rate (specificity), and true positive rate (sensitivity). In the context of this study, the negative class represents subjects who remain cognitively normal within a period of 10 years, while the positive class represents to subjects who develop MCI or AD within a period of 10 years. The metrics are calculated from the following equations:
\begin{equation}
\mathit{Accuracy = \frac{TN+TP}{TN + TP + FN + FP}}
\end{equation}
\\
\begin{equation}
\mathit{True \ Positive \ Rate = \frac{TP}{TP + FP}}
\end{equation}
\\
\begin{equation}
\mathit{True \ Negative \ Rate = \frac{TN}{TN + FN}}
\end{equation}
\section{Results and Discussion}
In this section, we present the results of our trained models. The comparison to state of the art is difficult due to the lack of studies which predict the progression of CN to MCI and AD. Furthermore, of the limited studies which do predict the conversion of CN individuals to MCI or AD, different datasets, subject selection criteria, and time to cognitive decline are utilised, making fair comparisons of results across studies difficult. As such, we produce our own baselines to measure model performance. To assess the validity of our hypothesis on attention-based mechanisms in the fusion of neuroimaging features, we first compare the results of MNA-net against a variant using no attention mechanisms. Finally, we perform an ablation study and investigate two aspects of the architecture: the efficacy of using patch-based techniques, and the value of using multimodal neuroimages in model performance. 

\subsection{Evaluation of Attention Based Mechanisms in Neuroimage Feature Fusion}
\begin{figure}[h]
    \includegraphics[width=0.45\textwidth]{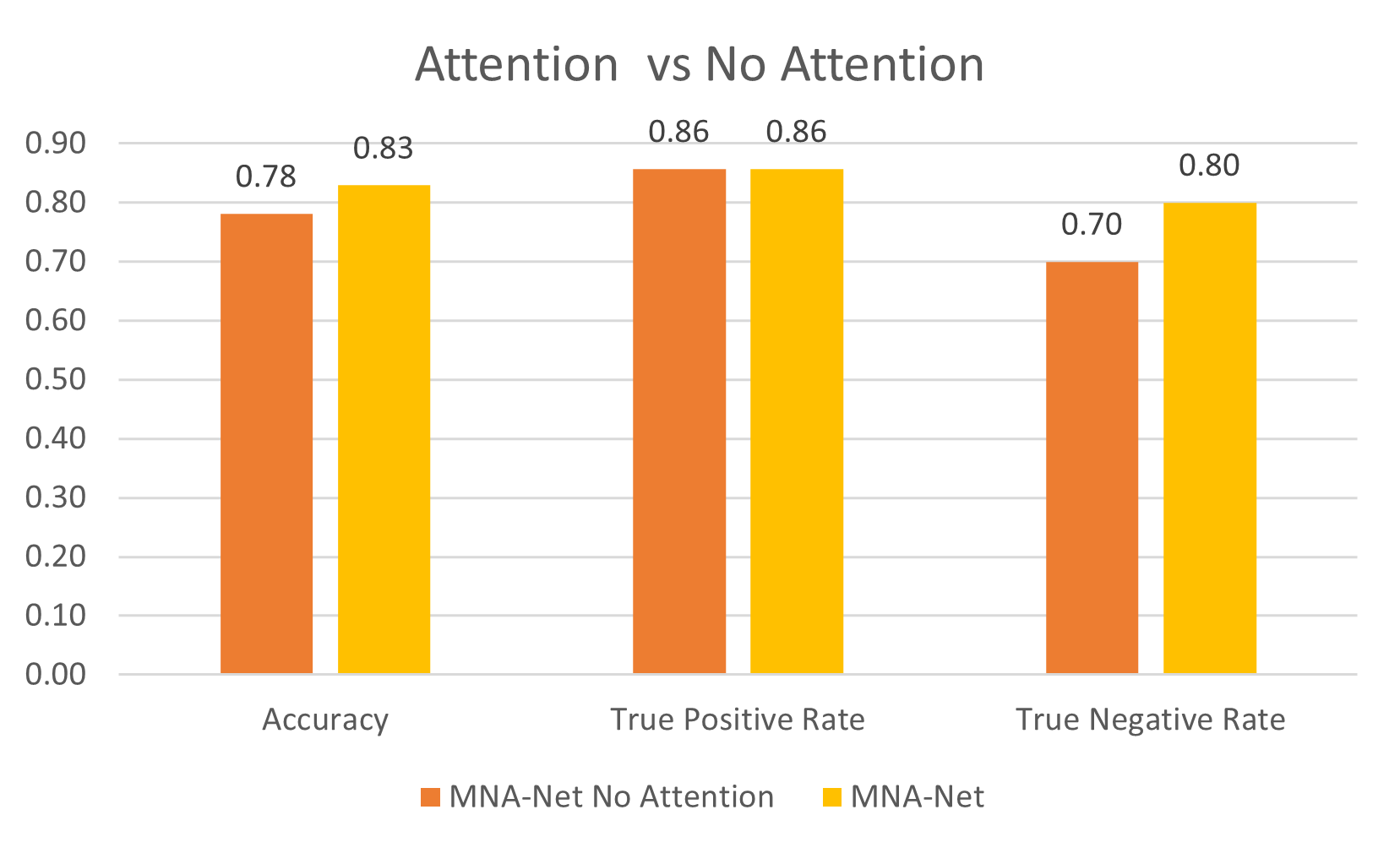}
    \centering
    \caption{Model performance of MNA-net: with and without attention mechanisms.}
    \label{fig:result1}
\end{figure}

To evaluate the impact of attention-based mechanisms in our proposed model, we train a variant of MNA-net which does not use attention-based mechanisms. For this variant, we replace the multi-head attention block with a dense layer during the multimodal attention stage. Classification performances for both the variants are presented in Figure \ref{fig:result1}. The results show that MNA-net achieves an increase in classification accuracy by 5\% and true negative rate by 10\%, while true positive rates remain the same when compared to its no attention variant.

The increase in accuracy and true negative rate in MNA-net suggests that incorporating attention based mechanisms improved the model’s ability in classifying individuals who remain cognitively normal over a 10 year period. This supports our hypothesis where the shared representations of MRI and PET features may provide more meaningful information and thus aid in the prediction of cognitive decline. Certain features in the PET images may provide contextual information to enhance certain features in the MRI images and vice versa. The concatenation of MRI and PET features seen in no attention variant of MNA-net limits the model’s ability to learn shared representations and therefore results in a lower performance.
\subsection{Ablation study}
\subsubsection{Evaluation of Patch-based Feature Extraction in classification performance}
\begin{figure}[h]
    \centering
    \includegraphics[width=0.5\textwidth]{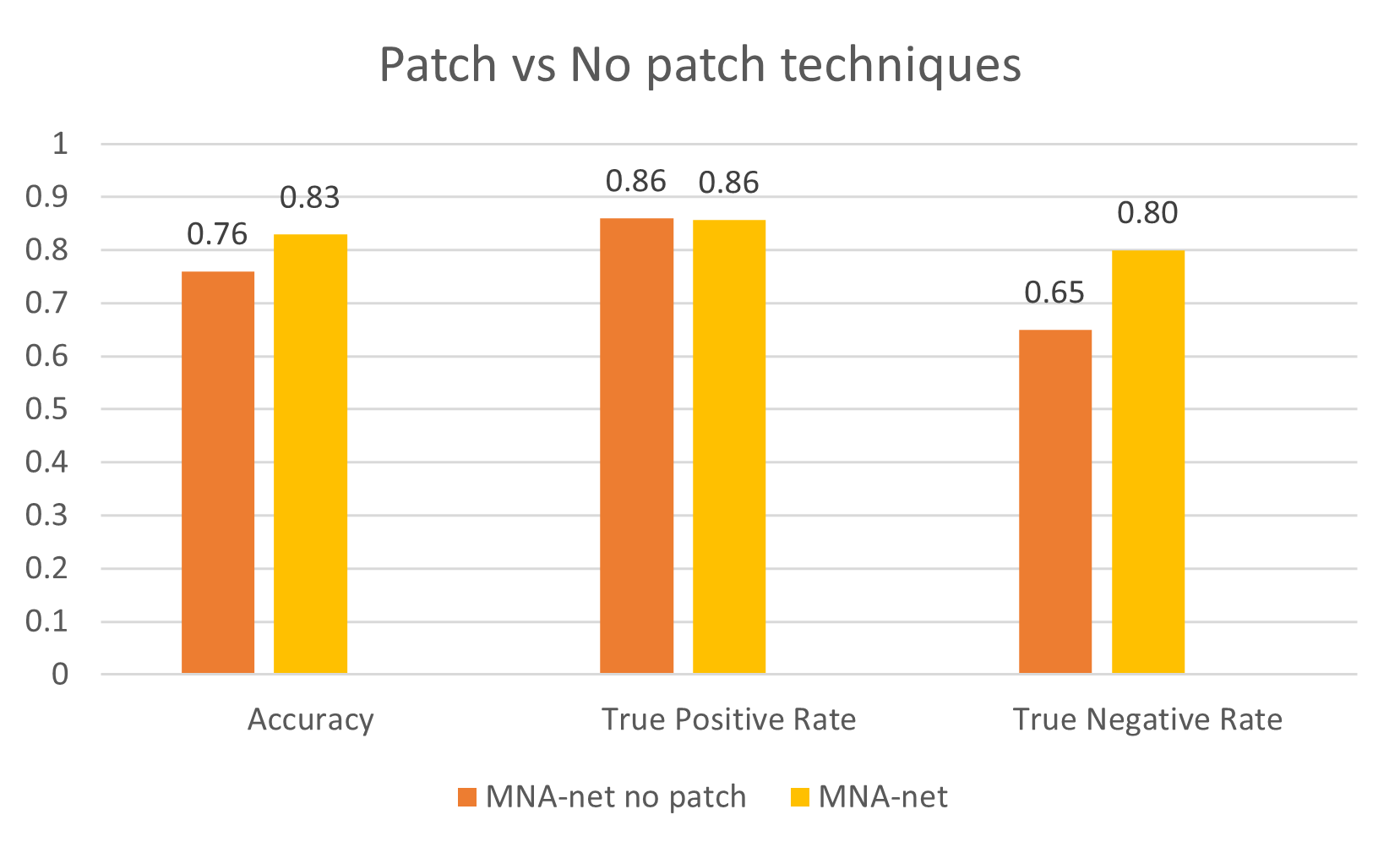}
    \caption{Comparison of the proposed model performance: MNA-net and MNA-net with no patch-based technique.}
    \label{fig:result2}
\end{figure}

To evaluate the importance of patch-based techniques, we compare MNA-net to a variant trained using no patch-based techniques. For this variant, instead of dividing neuroimages into 27 uniform patches, entire MRI and PET images are fed into two ResNet-10 models during the feature extraction stage. Classification performances of both the models are presented in Figure \ref{fig:result2}. Results demonstrate that MNA-net performance improves by 7\% and 15\% in accuracy and true negative rate, respectively, when compared to the variant trained with no patch-based techniques.

The superior performance of MNA-net can be explained due to the fact that different regions of the brain may have varying information in regards to the prediction of cognitive decline. For example, the atrophy in the hippocampus of the brain is commonly associated with AD \cite{johnson2012brain}. By dividing neuroimages into patches, the model is able to focus on smaller regions of the image and therefore localise and extract relevant features more effectively.

\subsubsection{Evaluation of Multimodal Neuroimages in classification performance}
To measure the importance of different modalities in MNA-net, we train three different models: a model trained on MRI images, a model trained on PET images, and a multimodal model trained on both neuroimaging modalities. For the unimodal models, we use the 3D ResNet-10 as shown in Figure \ref{fig:Resnet}. To have the true comparison among modalities, we modify MNA-net’s architecture for multimodal model by extracting features from the entire volumes instead of patches in the patch feature extraction phase, and substituting the multi-head attention block with a dense layer in the multimodal attention phase. 

\begin{figure}[h]
    \centering
    \includegraphics[width=0.5\textwidth]{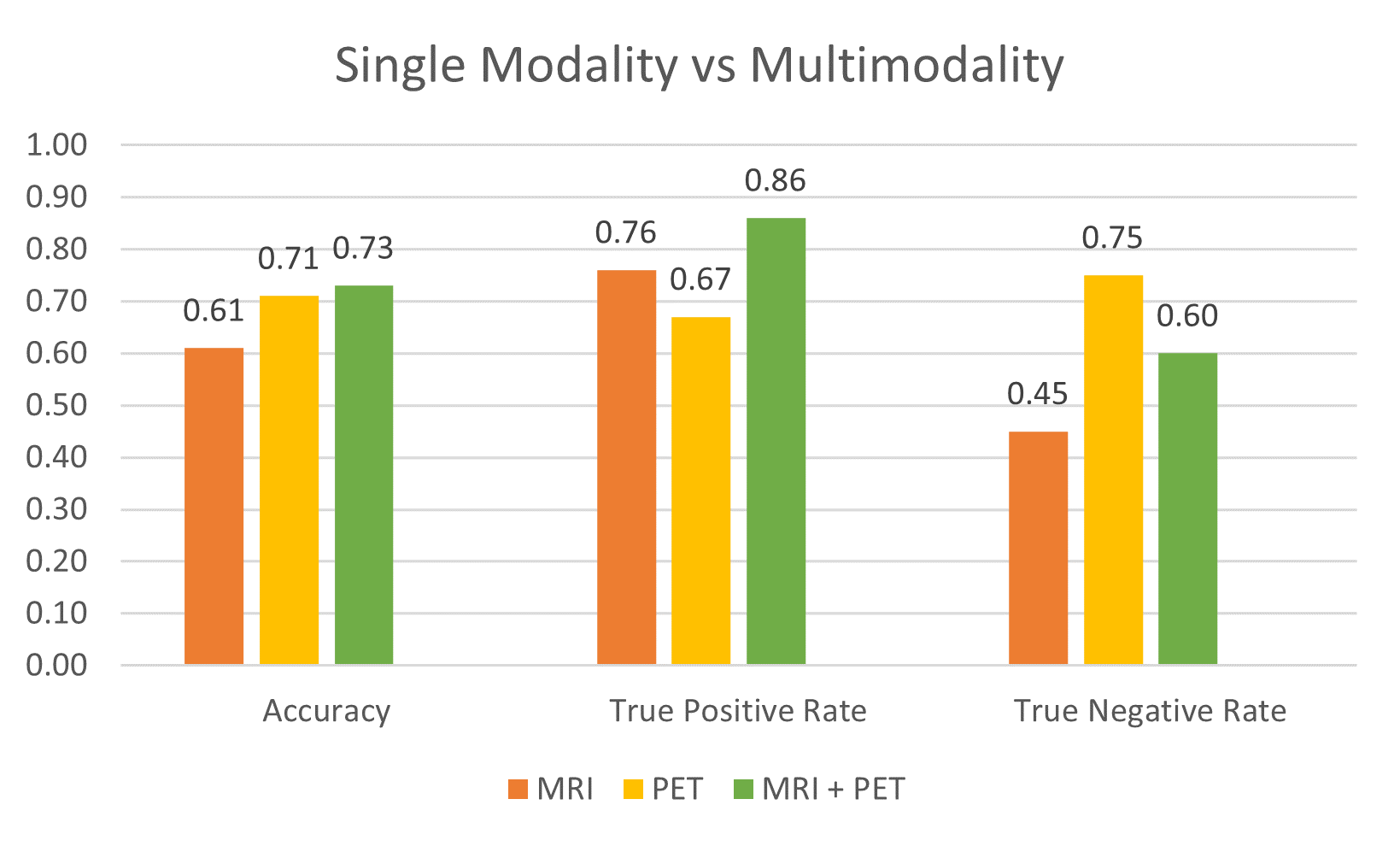}
    \caption{Model performance based on neuroimaging modality.}
    \label{fig:result3}
\end{figure}

Classification performances of all three models are shown in Figure \ref{fig:result3}. The model trained on both PET and MRI images was able to achieve the highest accuracy and true positive rate of 73\% and 86\% respectively. However, it exhibited a lower true negative rate compared to the model trained on PET. When comparing the single modalities, we can see that the use of PET images outperformed the MRI images, seeing increases of 10\% and 30\% in accuracy and true negative rate respectively. The true positive rate of the model trained on PET images however was slightly lower than that of the model trained on MRI images.

\begin{figure}[h]
    \centering
    \includegraphics[width=0.5\textwidth]{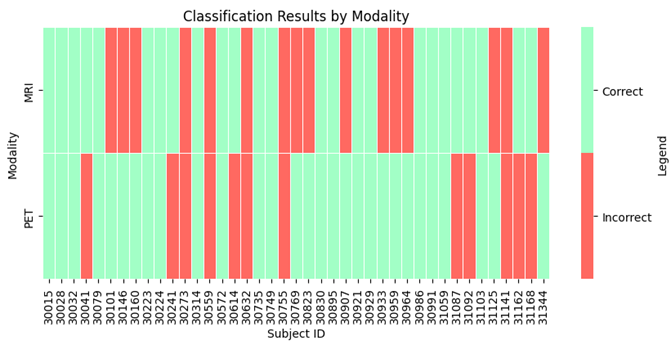}
    \caption{Proportion of correct and incorrect classifications by modality}
    \label{fig:result4}
\end{figure}

The superior performance of the multimodal model can be primarily attributed to the combination of different neuroimaging modalities. Both PET and MRI provide different information into the brain’s composition and structure, and therefore enabling the two neuroimaging modalities to correctly classify individuals who develop cognitive decline that the other may overlook. To confirm this, we investigate the proportion of correct and incorrect classifications of cognitive decline in subjects by neuroimaging modality. It can be observed in Figure \ref{fig:result4}, we can see that model trained on MRI images was able to successfully predict cognitive decline in seven subjects which the PET images could not, while the model trained on PET images was able to predict cognitive decline in eight subjects that the MRI model failed to recognise. Thus, the combination of PET and MRI modalities provide the model with more comprehensive information and features that complement each other to aid in the detection of cognitive decline. 

However, the possibility of incorporating poor or irrelevant features of each modality into the model poses as a challenge. For example, when examining the unimodal models, the model trained on the MRI images exhibited a low true negative rate of 45\%, while the model trained on PET images achieved a true negative rate of 75\%. The combination of the modalities in the multimodal model resulted in an intermediate true negative rate of 60\%, suggesting that the model may have retained some of the less relevant features from the MRI images, and thus resulting in its lower true negative rate compared to the PET model.

\subsection{Limitations}
Despite the promising results of MNA-net, the model is computationally intensive due to the wideness of the architecture. For example, for just only the patch extraction stage, 54 models in total are required to be trained (27 for each patch, for each modality). Thus, training the model is very memory intensive, computationally intensive and time consuming process. Furthermore, the OASIS-3 dataset after subject selection becomes relatively small, with sizes of only 122 for the training set without augmentation, and 41 for validation and test sets. Both these limitations combined makes MNA-net prone to overfitting.
The transfer learning could be implemented to overcome the limitation of small datasets and reduce computational requirements.

\section{Conclusion}
The combination of multiple neuroimages such as PET and MRI provide a more comprehensive view and understanding of the pathogenesis of MCI and AD. Therefore, more focus is observed in research into the applications of multimodal neuroimages in the classification of MCI and AD. However, in previous works, fusion of the different neuroimaging modalities simply involved the concatenation of learnt features, limiting their model’s ability to learn a shared representation and the complex relationships between each modality. Moreover, focus of research is found to be in detection of AD and MCI. In this work, we propose a novel multimodal neuroimaging attention-based CNN, MNA-net, to predict the conversion of CN individuals to MCI or AD in 10 years. We investigated the impact of attention-based mechanisms for the fusion of multimodal neuroimages in model performance. The experimental results demonstrate that the proposed MNA-net performs well and has increased accuracy and true negative rate due to attention mechanisms and provide a new state of the art results.
The results demonstrate that the shared representations of the PET and MRI images from the attention layers provided much more meaningful information. 
Furthermore, we are able to demonstrate the superiority of patch-based techniques and multimodal data in our proposed model's performance.

For future research, this work can be extended to investigate the use of attention based mechanism at the patch fusion level to improve model performance. Similar to the fusion of neuroimaging features, shared representations of different patch features, which consider features across all patches, may provide more meaningful information and therefore aid in prediction of cognitive decline.
In addition, further research into the use of multiple modalities may also be expanded to include non-imaging data 
such as clinical test scores and genetic data. 


%






%
\section*{Acknowledgements}
Data was provided by OASIS-3: Principal Investigators: T. Benzinger, D. Marcus, J. Morris; NIH P50 AG00561, P30 NS09857781, P01 AG026276, P01 AG003991, R01 AG043434, UL1 TR000448, R01 EB009352. AV-45 doses were provided by Avid Radiopharmaceuticals, a wholly owned subsidiary of Eli Lilly.

\section*{Declaration of competing interests}
The authors declare that they have no known competing financial interests or personal relationships that could have appeared to influence the work reported in this paper.
\bibliographystyle{ieeetr}
\bibliography{references_2.bib}




\end{document}